\documentclass{iaus}
\usepackage{graphicx}

  \checkfont{eurm10}
  \iffontfound
    \IfFileExists{upmath.sty}
      {\typeout{^^JFound AMS Euler Roman fonts on the system,
                   using the 'upmath' package.^^J}%
       \usepackage{upmath}}
      {\typeout{^^JFound AMS Euler Roman fonts on the system, but you
                   dont seem to have the}%
       \typeout{'upmath' package installed. iaus.cls can take advantage
                 of these fonts,^^Jif you use 'upmath' package.^^J}%
      }
  \else
  \fi

  \checkfont{msam10}
  \iffontfound
    \IfFileExists{amssymb.sty}
      {\typeout{^^JFound AMS Symbol fonts on the system, using the
                'amssymb' package.^^J}%
       \usepackage{amssymb}%
       \let\le=\leqslant  
       \let\ge=\geqslant  
      }{}
  \fi

  \IfFileExists{amsbsy.sty}
    {\typeout{^^JFound the 'amsbsy' package on the system, using it.^^J}%
     \usepackage{amsbsy}}
    {\providecommand\boldsymbol[1]{\mbox{\boldmath $##1$}}}

\newsavebox{\astrutbox}
\sbox{\astrutbox}{\rule[-5pt]{0pt}{20pt}}

\title[Large-Scale Environments of Groups \& Clusters]
      {The Large-Scale Environment of Groups \& Clusters of Galaxies}

\author[M.Plionis]
{Manolis Plionis$^{1,2}$}
\affiliation{$^1$ Institute of Astronomy \& Astrophysics, National Observatory
of Athens, P.Penteli, 152 36 Athens, Greece \\[\affilskip]
$^2$ Instituto Nacional de Astrof\'{\i}sica \'Optica y
Electr\'onica, AP 51 y 216, 72000, Puebla, Pue, M\'exico}

\pubyear{2004}
\volume{195}
\pagerange{1--8}
\date{?? and in revised form ??}
\jname{Outskirts of Galaxy Clusters: intense life in the suburbs}
\editors{A. Diaferio, ed.}

\begin{document}
\maketitle

\begin{abstract}
It appears that the dynamical status of clusters and groups of galaxies
is related to the large-scale structure of the Universe.
A few interesting trends have been established:

\noindent
(1) {\em The Cluster Substructure - Alignment Connection} by which clusters 
show a strong correlation between their tendency to be aligned with
their neighbors and their dynamical state (as indicated by the existence of
significant substructres). 

\noindent
(2) {\em The Cluster Dynamics -Cluster Clustering Connection} 
by which dynamically 
young clusters are more clustered than the overall cluster population.

\noindent
(3) {\em The Cluster- Supercluster Alignment Connection} by which
clusters of galaxies show a statistical significant tendency to be aligned 
with the projected major axis orientation of their parent supercluster.

\noindent
(4) {\em The Galaxy Alignment - Cluster Dynamics Connection}
by which red-sequence cluster bright galaxies 
show a significant trend to be aligned
with their parent cluster major axis, especially in dynamically young clusters.

\noindent
(5) {\em The Group Richness - Shape Connection} by which groups of
    galaxies are flatter the poorer they are.

These are strong indications that clusters develop in a hierarchical 
fashion by anisotropic merging of smaller units along the large-scale 
filamentary structures within which they are embedded.
\end{abstract}

\section{Introduction}
In the framework of the hierarchical model for the formation of cosmic 
structures, galaxy clusters are supposed to form by accretion of
smaller units (galaxies, groups etc).
After the epoch of mass aggregation (which depend on the
cosmological model), violent relaxation processes will tend to
alter the cluster phase-space configuration 
producing `regular', quasi-spherical, having
smooth density profile clusters.

In the last decade due to the increased spatial resolution in X-ray 
imaging ({\sc Rosat}, {\sc Xmm}-{\em Newton}, {\sc Chandra}) 
and to the availability of wide-field cameras, many of the previously 
thought ``regular'' clusters have shown to be clumpy to some level, 
a fact that could have important consequences for structure 
formation theories, since the present fraction of dynamically young
clusters, as well as the rate of cluster evolution (as measured for
example by their luminosity and temperature functions and their
morphology), are cosmology dependent
(eg., Richstone, Loeb \& Turner 1992, Evrard et al. 1993, 
Lacey \& Cole 1996).
Detailed optical and X-ray studies have yielded evidence
for significant amounts of subclustering within local and distant 
clusters (eg. Buote \& Tsai 1996; Plionis \& Basilakos 2002; 
Jeltema et al. 2004) 
indicating that possibly a large fraction is still forming at the present time.
Hints do exist for a very recent (within the last Gyr) dynamical 
evolution of the cluster population (Melott et al. 2001;
Plionis 2002, see however Floor et al. 2003). 
Note however that the existence of substructure does not necessarily mean that
the corresponding cluster is dynamically young due to the ambiguity of cluster
post-merging relaxation times (see discussion in Plionis 2001).

Furthermore, the dynamical evolution of member galaxies and of the ICM
gas is also an open issue. Star formation seems to be active in
clusters showing substantial substructure and velocity
gradients, as expected if a recent merger has taken place. For
example, the fraction of blue galaxies is strongly correlated with
cluster ellipticity (Wang \& Ulmer 1997), while ellipticity
is strongly correlated with the dynamical state of the cluster 
(eg. Kolokotronis et al. 2001). It appears
that the violent merging events trigger star-formation, possibly through a
multitude of different mechanisms; for example, the excess number of 
galaxy-galaxy interactions, the rapid variation of the cluster 
gravitational field (Bekki 1999), etc. 
An interesting observable that may be related to the 
dynamics of clusters
is their tendency to be aligned with their nearest 
neighbor as well as with other clusters that reside in the same 
supercluster (eg. Bingelli 1982; Plionis 1994). 
Analytical (Bond 1986) and numerical work (eg.
West et al. 1991, van Haarlem \& van de Weygaert 1993, 
Splinter et al. 1999, Onuora \& Thomas 2000;
Faltenbacher et al. 2002; Knebe et al. 2004)
have shown that such alignments occur in many hierarchical clustering
models and are probably the result of an interesting 
property of Gaussian random fields which is the "cross-talk" between 
density fluctuations on different scales (eg. West 1994).

\section{Environmental Trends}
Our aim is to investigate whether there is any relation between the 
cluster and group large-scale environment and their internal dynamical
state. To
this end we have analysed {\sc Apm} and Abell clusters (in total more than
1200 clusters) as well as the {\sc Uzc-Ssrs2} (Ramella et al. 2002) and
{\sc 2dfgrs} (Eke et al. 2004) samples of groups of galaxies. 
\begin{figure}
\includegraphics[width=7cm]{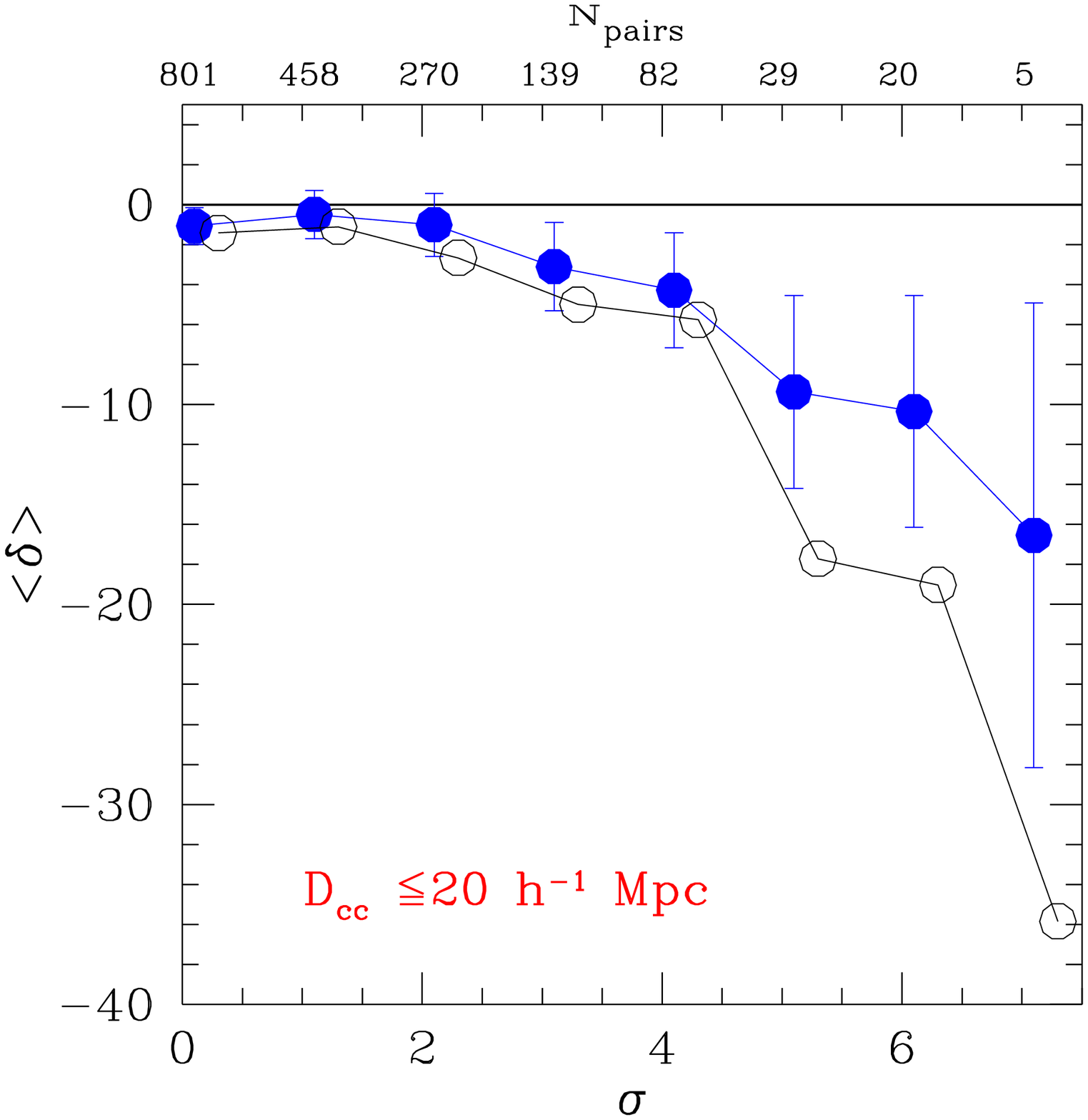}
\hfill
\includegraphics[width=7cm]{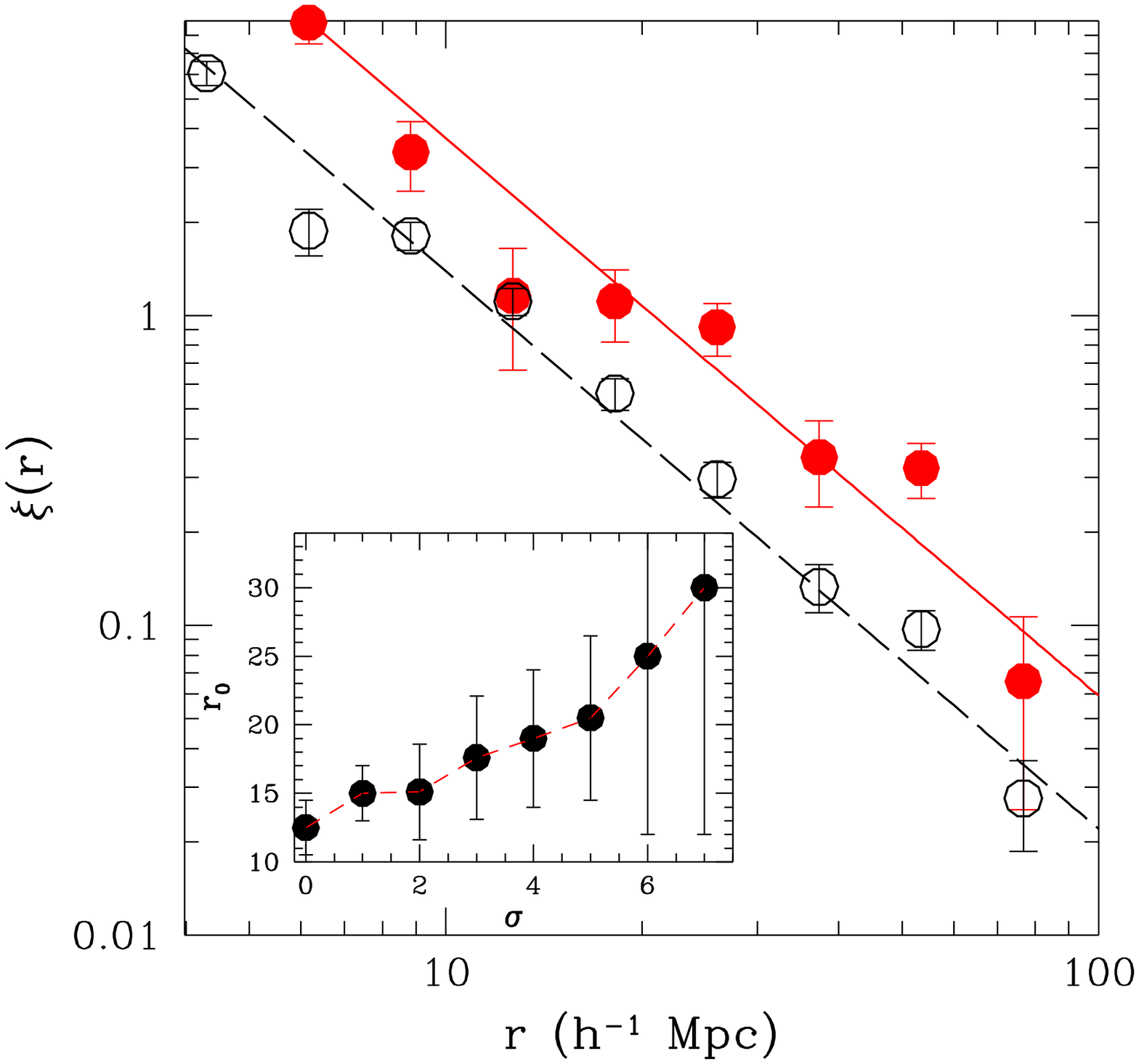}
\caption{\small
{\sc Left Panel}: Alignment signal as a function of substructure 
significance level. 
Filled circles correspond to nearest-neighbors and open circles 
to all neighbors within 20 $h^{-1}$ Mpc. 
{\sc Right Panel}: The {\sc Apm} cluster correlation function for 
all the {\sc Apm} clusters (open symbols) and for a sample with 
significant substructure (filled symbols). The lines
represent the best $(r/r_{\circ})^{-1.8}$ fit with $r_{\circ}\simeq12$ and
$\simeq 20$ $h^{-1}$ Mpc respectively. In the insert we show the correlation
length as a function of substructure significance level.}
\end{figure}

In order to realize our study we need to define in an objective manner
the dynamical state of the clusters. 
Evrard et al. (1993) and Mohr et al. (1995) have suggested as an 
indicator of cluster substructure the {\em centroid-shift} ($sc$)
which is defined as the shift of the cluster center-of-mass position 
as a function of density threshold above which it is estimated.
The significance, $\sigma$, of such centroid variations to
the presence of background contamination, random density
fluctuations and small galaxy number effects can be estimated using 
Monte Carlo cluster simulations in which, by construction, there is no
substructure (see Plionis 2001).
Kolokotronis et al. (2001) calibrated various substructure measures
using {\sc Apm} galaxy data and {\sc Rosat} pointed observations for 
22 Abell clusters
and found that in most cases using either X--ray or optical data one can
identify substructure unambiguously.  
Their conclusion was that a large and significant {\em centroid-shift}
is a clear indication of cluster substructure.
\begin{figure}
\includegraphics[width=6.5cm]{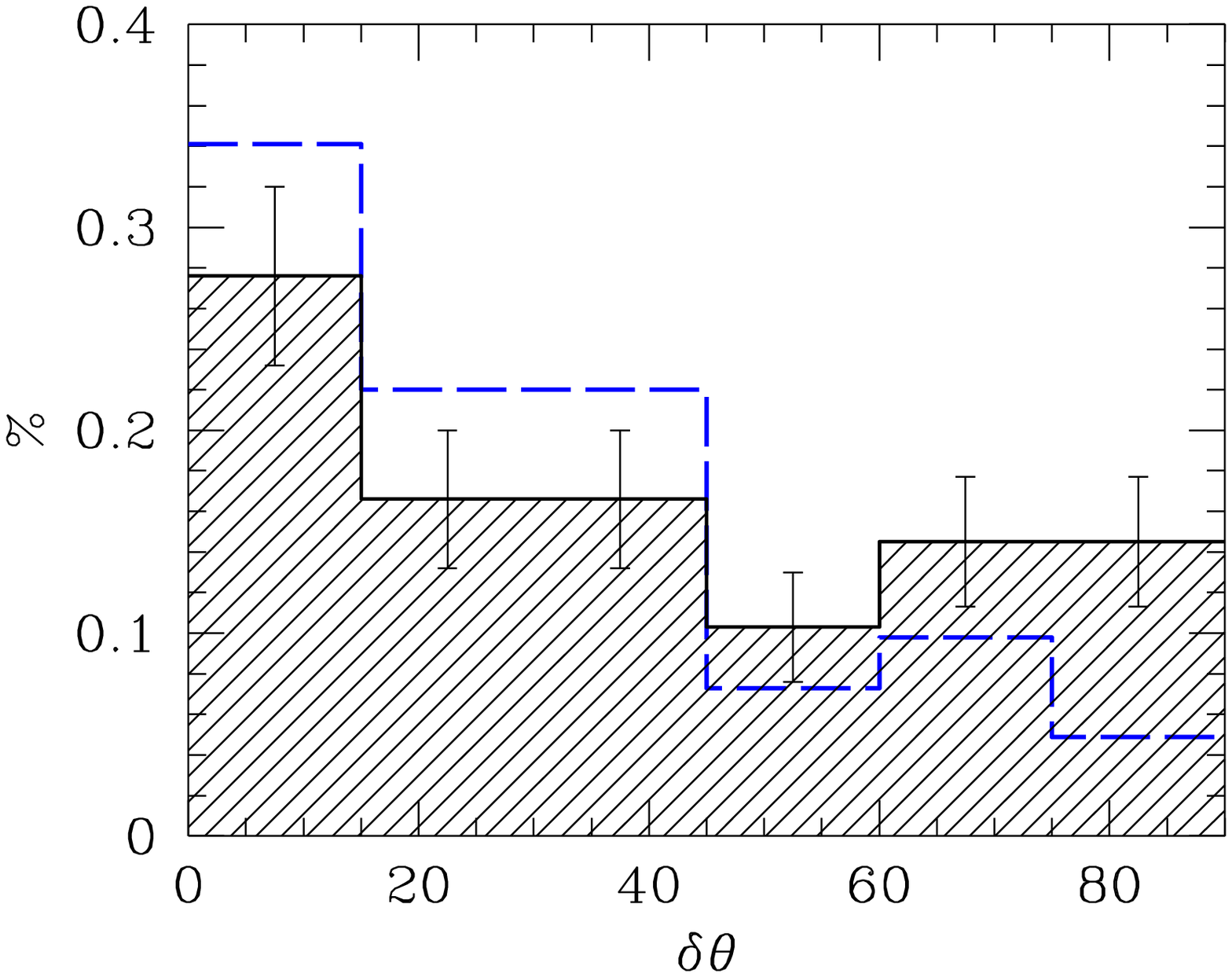}
\hfill
\includegraphics[width=6.5cm]{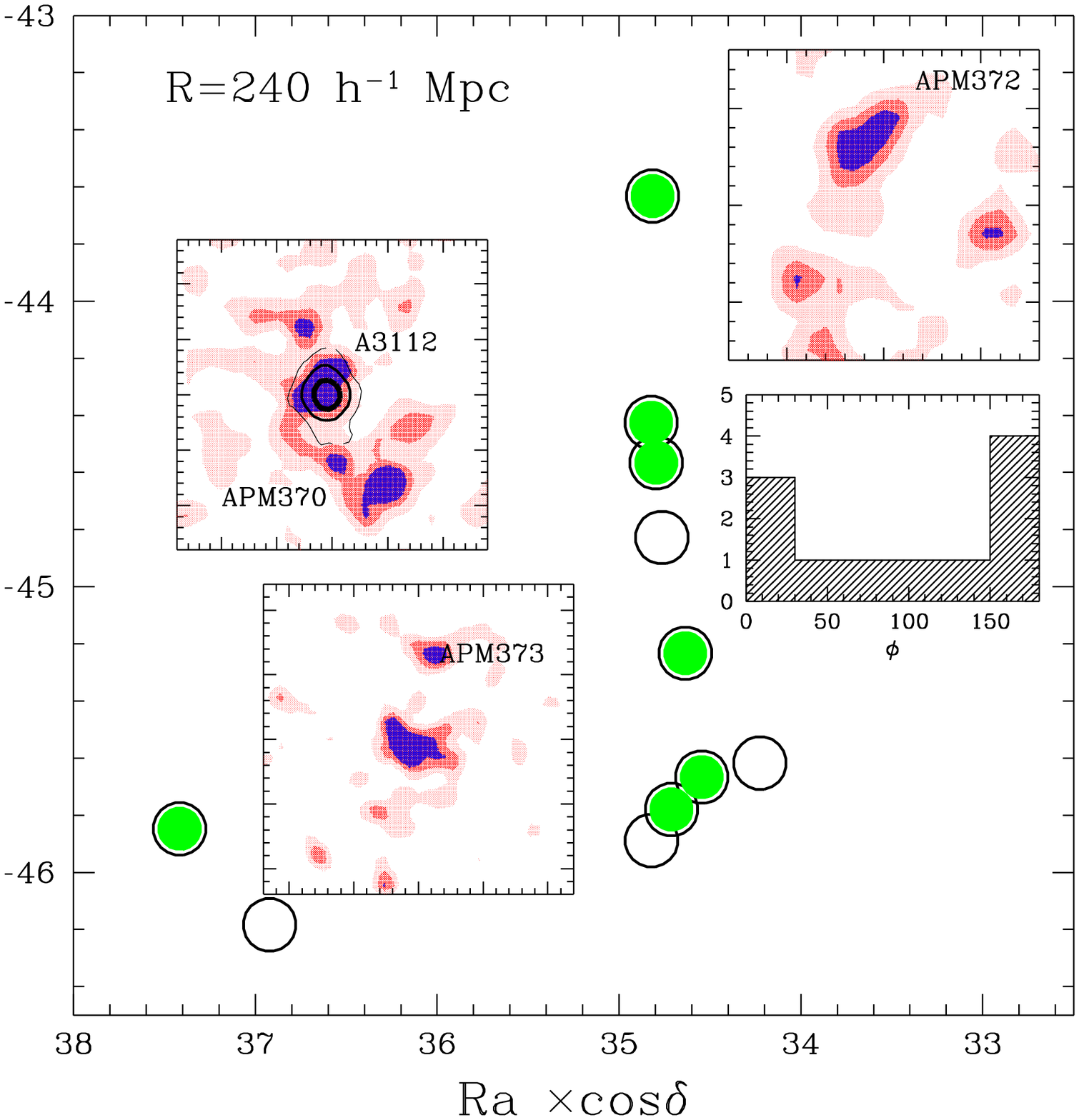}
\caption{\small {\sc Left panel:} 
Frequency distribution of the misalignment angle between
cluster members and their parent supercluster orientations.
The dashed line corresponds to superclusters defined with linking radius of
30 $h^{-1}$ Mpc while the hatched distribution to that with linking
radius of 20 $h^{-1}$ Mpc.
{\sc Right panel:}
An example of a filamentary {\sc Apm} supercluster containing A3112, A3104,
A3111 as well as 9 poorer {\sc Apm} clusters, identified using a linking radius
of 12 $h^{-1}$ Mpc. Filled dots represent clusters with significant 
substructure. In the inserts we show the smooth density distribution
for three of the clusters (for the case of A3112 we also show its
{\sc Rosat-Pspc} X-ray contours) as well as the frequency distribution
of cluster
position angles. The preferred cluster major axis 
alignment with the projected orientation of the supercluster is evident.}
\end{figure}

Furthermore, in order to investigate the alignment between cluster, 
group or galaxy orientations,
we use the relative position angle between pairs, $\phi_{i,j}$.
A significant deviation from the isotropic expectation
can be quantified by the measure:
$\delta=\sum \phi_{i,j}/N-45$ (Struble \& Peebles 1995) which for
an isotropic distribution it has a mean value $\langle \delta \rangle
\simeq 0$, and a standard deviation given by $\sigma=90/\sqrt{12 N}$. 
A significantly negative value of $\delta$ would indicate alignment while
a positive misalignment.

Below we
present the main results of a variety of studies that indicate a
strong correlation between the dynamical state of
galaxy structures (groups and clusters) and
their large-scale environment.

\subsection{Cluster Substructure - Alignment Connection}
Plionis \& Basilakos (2002), using a sample of $\sim 900$ {\sc Apm}
 clusters, correlated the alignment signal
between cluster neighbors, $\langle \delta \rangle$, with 
the existence of cluster substructure, as measured by the deviation
 from the expected density fluctuations in units of $\sigma$
 (eg. Kolokotronis et al 2001; Plionis 2001).
In the left panel of figure 1 we present $\langle \delta \rangle$
between cluster nearest-neighbors (filled dots) and between all pairs
(open dots) with pair separations $< 20$ $h^{-1}$ Mpc. 
Evidently, there is a strong correlation
between the strength of the alignment signal and the substructure
significance level. 
This result confirms statistically the analysis of West et al.
(1995) based on {\sc Einstein} X-ray data of $\sim 90$ clusters.

\subsection{Cluster Substructure - Clustering Connection}
Where do clusters with significant substructure reside?
The spatial 2-point cluster correlation function for cluster samples 
with different
substructure significance levels show a strong trend.
In the right panel of figure 1 we plot the correlation length,
$r_{\circ}$, as a
function of cluster substructure significance level, which is 
clearly an increasing function of $\sigma$.
The conclusion of this analysis is that clusters
showing evidence of dynamical activity reside in high-density
environments. Our results are in agreement with a similar 
environmental dependence found in
the {\sc Reflex} and {\sc Bcs} 
X-ray cluster sample (Sch\"{u}ecker et al., 2001) and
for cooling flow clusters with high mass accretion rates (Loken, Melott
\& Miller 1999).
\begin{figure}
\includegraphics[width=7cm]{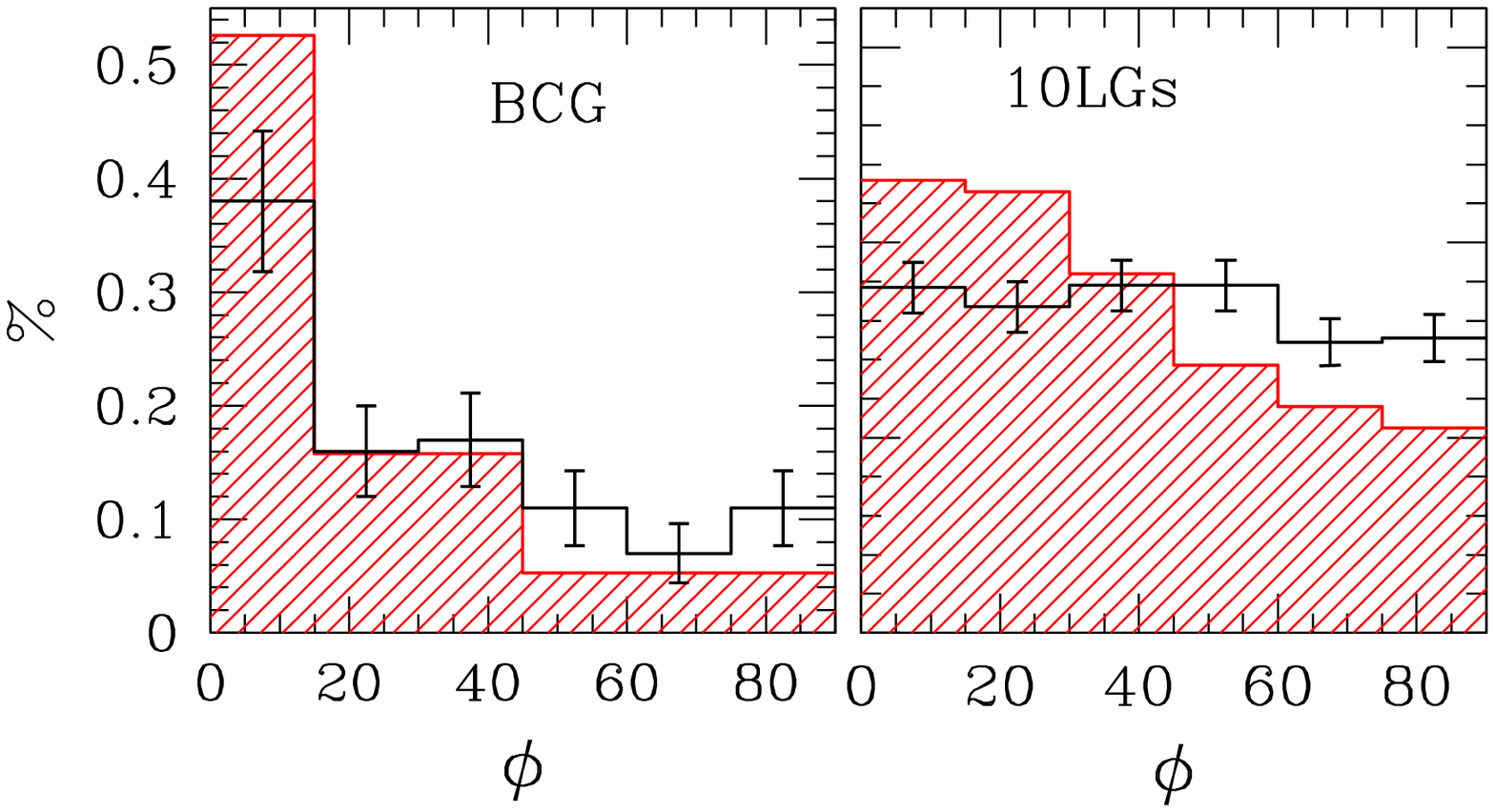}
\hfill
\includegraphics[width=7cm]{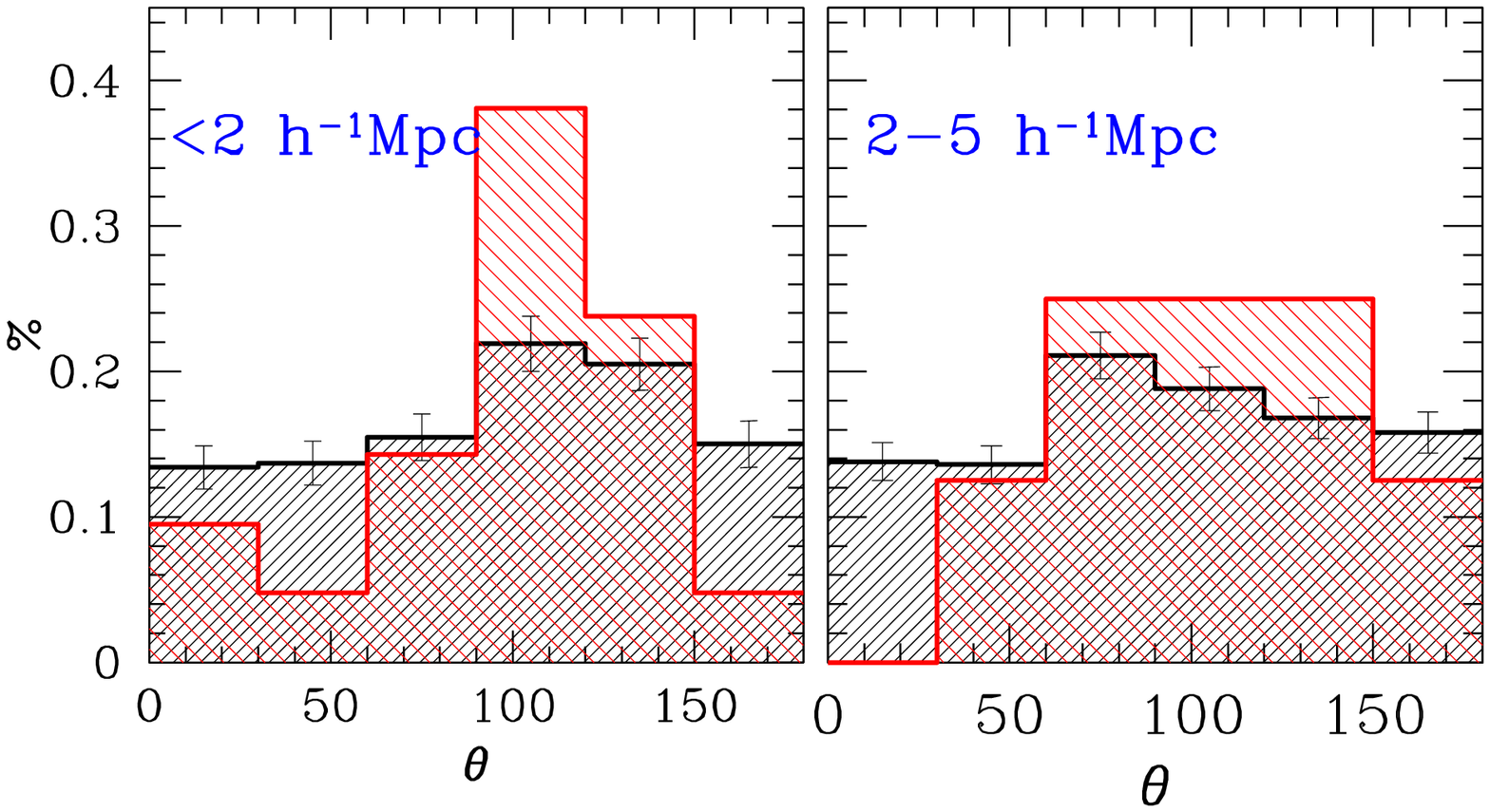}
\caption{\small {\sc Left panel:} 
The distribution of misalignment angles $\phi$
between the galaxy and cluster orientations within superclusters
for the $v_{\sigma}<900$ km/sec 
clusters (black lines) and for those with $v_{\sigma} \ge 900$ km/sec
(red hatched histogram).
{\sc Right panel:} Histogram of galaxy (black lines) and group
(red lines) position angle distribution in Abell 521 for two 
indicated distance shells.}
\end{figure}

\subsection{Cluster - Supercluster Alignment Connection}
A further interesting question regarding environmental effects on
large scales
is whether clusters are also aligned with the orientation 
of their parent supercluster. 
We define {\sc Apm} superclusters by using
a friends of friends algorithm and estimate for each supercluster
the misalignment angle, $\delta\theta$, between its projected orientation
and the mean position angle of its member clusters. In figure 2 we
present the frequency distribution of $\delta\theta$ for two different
supercluster catalogues (based on linking radii of 20 and 30
$h^{-1}$ Mpc, respectively). The significant excess of small
$\delta\theta$'s is evidence that indeed clusters do
show preferential alignments with the orientation of their parent
superclusters. As an example, in the right panel of Fig.2 we present 
the projected distribution of clusters in a dense and filamentary 
{\sc Apm} supercluster.
Embedded in the plot are the smoothed galaxy density distributions of a few
member clusters. It is evident that
the cluster position angle distribution reflects the projected
orientation of the supercluster.

\subsection{The Galaxy Alignment - Cluster Dynamics Connection}
Analysing the DPOSS images of 303 Abell clusters (with $z\le 0.12$),
Plionis et al. (2003) found significant alignments 
between the projected major axis of the 10 brightest galaxies 
(excluding the BCG) and their parent cluster position angle, mostly
in high-density environments. This 
alignment signal is stronger for higher velocity dispersion clusters,
which in our case (based on a variety of criteria) 
appear also to be dynamically young merging clusters (see left panel
of Fig.3). This
points strongly in the direction of galaxy alignments being 
correlated with their parent cluster dynamical youth and local density
(see also West \& Blakeslee 2000; Knebe et al. 2004).
We have also studied in detail, using very deep WFI images, the
highly unrelaxed Abell cluster (A521 at $z=0.25$) and find that such alignments
(see right panel of Fig.3)
are present also in fainter red-sequence galaxies as well as in the
orientation of groups of galaxies in the extended environment around
the cluster (see also Kitzbichler \& Saurer 2003).
\begin{figure}
\center
\includegraphics[width=11cm]{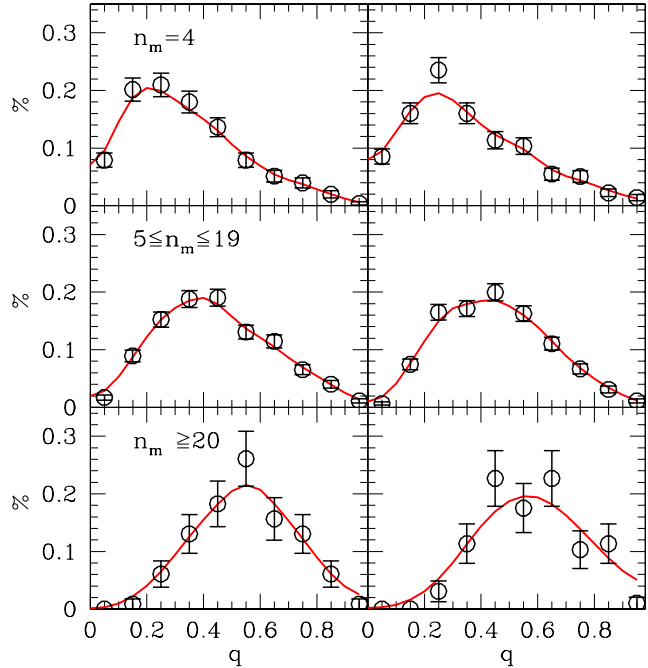}
\caption{\small 
The apparent {\sc 2dfgrs} group axial ratio
distributions for different group membership, $n_m$, and for the 
northern (NGP) and southern (SGP) subsamples respectively 
The solid line is the smooth fit from the nonparametric kernel 
estimator.}
\end{figure}

\subsection{Group Shape - Richness Connection}
Plionis, Basilakos \& Tovmassian (2004) and Plionis \& Basilakos
(2004) analysed two large samples of groups (the {\sc Uzc-Ssrs2} 
and the {\sc 2dfgrs-2pigg} samples) within a
roughly volume limited region ($cz\le 5500$ and $cz\le 30000$ km/sec
for the two samples respectively). 
In figure 4 we present the projected, on 
the plane of the sky, axial ratio distribution for three group
richness classes, while in figure 5
we show the corresponding inverted
axial ratio distribution assuming that groups are intrinsically
either prolate or oblate spheroids. 

There is an obvious group richness-flatness relation,
seen in both projected and intrinsic axial ratio distributions,
with group flatness decreasing with richness. 
We also find a weak (Spearman correlation coefficient ${\cal R}=-0.13$)
but significant correlation (${\cal P}<10^{-6}$), 
between the velocity dispersion of groups and their flattening, with
flatter systems having lower velocity dispersions.
Such a correlation as well as 
the increase of the group sphericity with richness, which extends also
to clusters, could be explained as an indication of a higher degree of
virialization, which is expected to be more rapid in more massive
systems. Furthermore, we find that groups of galaxies are prolate-like
systems which is agreement with the shape of clusters and
superclusters of galaxies (eg. Cooray 2000; Basilakos, Plionis \&
Maddox 2000 and
references therein). This is the expected shape for cosmic
structures that form in quasi one-dimensional large-scale structures 
(filaments). Other interesting environmental effects that 
have been found is the decrease of the 
group velocity dispersion as their distance from nearby clusters decreases
(eg. Ragone et al 2004).

\begin{figure}
\center
\includegraphics[width=11cm]{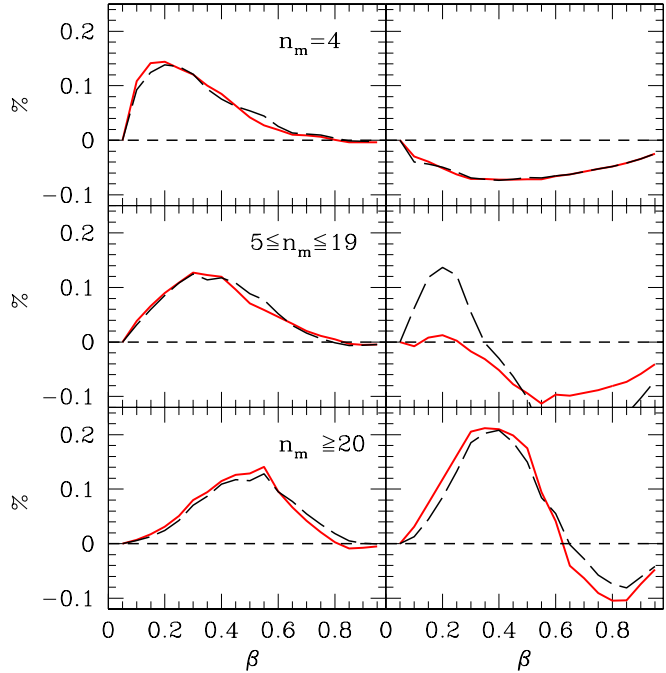}
\caption{\small
{\sc Right panel:} The intrinsic distribution of group axial ratios
for the NGP (continuous line) and SGP (dashed line) subsamples
assuming that they are either prolate (left panel) or oblate (right
panel) spheroids. The unphysical negative values of the latter model
indicates its failure to represent the true three-dimensional shape
of groups.}
\end{figure}

\section{Conclusions}
We have presented evidence, based on the {\sc Apm} and Abell cluster samples
as well as the {\sc Uzc-Sssrs2} and {\sc 2dfgrs} group samples, 
that there is a strong link between the 
dynamical state of clusters/groups of galaxies 
and their large-scale environment.
Dynamically young clusters are significantly more aligned with their nearest
neighbors and they are also much more spatially clustered.
Clusters belonging in dense superclusters are preferentially aligned
with their parent supercluster projected orientation and 
bright galaxies, in dynamically young clusters, are also orientated
along their parent cluster major axis. These coherent orientation effects, from
the scale of galaxies to that of superclusters
support the hierarchical formation scenario in which clusters
form by anisotropic merging along the large-scale filamentary structures within
which they are embedded.

\acknowledgments
This work has been partly funded
within the framework of the program 'Promotion
of Excellence in Technological Development and Research', project
{\em 'X-ray Astrophysics with ESA's mission XMM'}, as well as
by the Mexican Government grant No CONACyT-2002-C01-39679.

\end{document}